\documentclass[prb,a4paper,showkeys,reprint]{revtex4-1}
\usepackage{graphics}
\usepackage{bm}
\usepackage{amsmath}
\usepackage{amssymb}
\usepackage{booktabs}
\usepackage{bbold}
\usepackage{color,xcolor}
\newcommand{\Ignore}[1]{}
\newcommand{\Allow}[1]{#1}

\begin{document}

\Allow{
\title{Spatial dispersion effects upon local excitation of extrinsic plasmons in a graphene micro-disk}
\author{D. Mencarelli$^{1,2}$, S. Bellucci$^{2}$, A. Sindona$^{2,3}$, L. Pierantoni$^{1,2}$}
\affiliation{
$^1$Universit\`{a} Politecnica delle Marche, Ancona, Italy\\
$^2$Laboratori Nazionali di Frascati (LNF--INFN), Frascati, Roma, Italy\\
$^3$Dipartimento di  Fisica, Universit\`{a} della Calabria, Via P. Bucci Cubo~30C, Rende (CS), Italy
}
\begin{abstract}
Excitation of surface plasmon waves in extrinsic graphene is studied using a full-wave electromagnetic field solver as analysis engine.
Particular emphasis is placed on the role played by spatial dispersion due to the finite size of the two-dimensional material at the micro-scale.
A simple instructive set up is considered where the near field of a wire antenna is held at sub-micrometric distance from a disk-shaped graphene patch.
The key-input of the simulation is the graphene conductivity tensor at terahertz frequencies, being modeled by the Boltzmann transport equation for the valence and conduction electrons at the Dirac points~(where a linear wave-vector dependence of the band energies is assumed).
The conductivity equation is worked out in different levels of approximations, based on the relaxation time ansatz with an additional constraint for particle number conservation.
Both drift and diffusion currents are shown to significantly contribute to the spatially dispersive anisotropic features of micro-scale graphene.
More generally, spatial dispersion effects are predicted to influence not only plasmon propagation free of external sources, but also typical scanning probe microscopy configurations.
The paper set the focus on plasmon excitation phenomena induced by near field probes, being a central issue for the design of optical devices and photonic circuits.

\end{abstract}
\maketitle
}

\section{Introduction}

\label{sec:intro} One of the main challenges for current device electronics
is the full exploitation of the terahertz~(THz) electromagnetic~(EM)
spectrum, bridging the gap beitween the microwaves and optics. In this
context, graphene and graphene-derived materials possess a number of unique
EM properties such as tunable conductivity and slow-wave features that may
be used to develop high performance THz devices~\cite{1,2,3,4,5,5b,6}. More
generally, low-dimensional systems with honeycomb-like geometry and related
assemblies or hetero-structure are undergoing massive investigation as
materials capable of supporting plasmon propagation in the THz frequency
range~\cite{7,7b,7c,7d,8b}.

A THz beam impinging on a metal atomic force microscope tip has been used to
generate guided THz waves on graphene in a recent experiment~\cite{7}: the
nanometric curvature of the tip makes light scatter on the length scale of
tens of nanometers~\cite{2}, providing the EM field together with the
wave-number components required to excite short wavelength surface plasmons.
A similar situation has been encountered at lower frequencies in scanning
microwave microscopy~(SMM)~\cite{8}, where a metallic tip can act as a
resonant antenna whose tip-termination is directly coupled with the sample.

In this paper, we focus on numerical propagation and excitation of plasmon
surface waves in spatially dispersive graphene as seen from the macroscopic
EM perspective. We use the fact that a momentum matching between the field
source and the surface plasmon can be achieved, e.g., by grating or prism
couplers. In particular, we present an EM scheme that ideally recalls the
set-up of Ref.~\cite{8} and take a resonant wire antenna placed at
sub-micrometric distance from a disk-shaped graphene patch.

Our analysis suggests that the correction to the conductivity response given
by spatial dispersion is strongly required, at least in excitation phenomena
induced by a strong spatial gradient of the EM field on the sample. We
therefore asses that spatial dispersion is an important limiting factor to
the achievable coupling between surface-plasmons and the near field of an
SMM tip, because it produces a broadening of the system-response over a wide
range of wave-vectors. The high impact of spatial dispersion in propagation
and modal analysis of plasmons has been previously scrutinized~\cite{9,10,11}%
. However, a similar study has not been conducted so far in near field
excitation problems, which are of central importance in numerous practical
applications including microscopy and photonics. Here, we provide such a
characterization.

From the numerical point of view, we simulate surface plasmon excitation in
an open computational domain avoiding periodic boundary conditions and
plane-wave excitations~\cite{5}. The typical high charge mobility of carbon
nanostructures leads to a huge increase of the macroscopic inductance, i.e.,
a high reactive energy stored per unit length and, accordingly, a slow-wave
propagation. Consequently, the plasmon velocity can be much smaller than the
speed of light with very small wavelengths as compared to that of the
coupled feeding antenna. This also results in a high confinement of the EM
field in the direction transverse to the propagation, which further
increases the aspect ratio. We then apply two different numerical methods to
validate the results, i.e., the method of moments~(MoM) and the method of
finite elements~(FEM). In this way, we provide a comparison of the different
levels of approximation used in semi-classical approaches to the graphene
conductivity, based on the Boltzmann transport equation for the Dirac
electrons and the Kubo-formula~\cite{12b,12c,12cc,12,12o}.

\section{Theory}

\label{sec:theory} The dispersive conductivity of graphene on the THz regime
is well defined in the literature~\cite{1,10,11}: basically, an electric
field frequency below a few THz induces a response of the valence~($\pi$)
and conduction~($\pi^{\ast}$) electrons of the material with an energy close
to the Fermi energy. The dominant part of the response in extrinsic~(doped
or gated) graphene is given by intra-band excitations, which can be treated
by the Boltzmann transport equation under the
relaxation-time-approximation~(RTA) or Bhatnagar-Gross-Krook~(BGK) model.

The RTA replaces the relaxation dynamics of each one-electron state by a
simple exponential decay. Then, the transport relaxation time is
approximated with the lifetime of the state. Such a formalism provides an
accurate description of spatial dispersion effects in doped graphene with
extrinsic Fermi energy shifts below $\sim 0.5$~eV. Accordingly, the RTA
conductivity is entirely determined by the drift currents that arise from
one-electron transitions and collective modes within the $\pi $ and $%
\pi^{\ast }$ bands.

The BGK model is more general than the RTA, because it allows for an extra
degree of freedom, which enforces charge conservation and accounts properly
for electron diffusion~\cite{10,11}.

We begin by specializing to the RTA conductivity
\begin{equation*}
\bm{\sigma}_{\text{\textsc{rta}}}(\mathbf{q})= \bm{\sigma}_{\text{\textsc{rta%
}}}^{+}(\mathbf{q})+\bm{\sigma}_{\text{\textsc{rta}}}^{-}(\mathbf{q})
\end{equation*}
under an applied electric field with momentum $\mathbf{q}$ and frequency $f$%
. $\bm{\sigma}_{\text{\textsc{rta}}}$ includes the contributions of
electrons~(with charge $-e$ and wave-vector $\mathbf{k}$) that occupy the $%
\pi^{*}$(+) and $\pi$(-) bands. The band energies $\varepsilon_{\pm}(k)$ are
populated according to the Fermi Distribution~(FD)
\begin{equation*}
{f}_{\varepsilon_{\pm}(k) -\mu }=\frac{1}{1+e^{\frac{\varepsilon_{\pm}(k)
-\mu }{kT}}}
\end{equation*}
at the absolute temperature $T$ and chemical potential $\mu$. We adopt the
widely used convention of setting the Dirac point energy of graphene, i.e.
the intrinsic Fermi level, to zero energy. In this way, $\mu$ coincides with
the Fermi level shift caused by the doping~(or gating).

The intra-band components of $\bm{\sigma}_{\text{\textsc{rta}}}$ have the
dyadic~(tensor) form~\cite{10}
\begin{equation}
\bm{\sigma}_{\text{\textsc{rta}}}^{\pm}(\mathbf{q})=\frac{ i e^2}{2 \pi^2}
\int\limits_{1^{\mathrm{st}}~\mathrm{BZ}} d^2 k \frac{f_{\varepsilon_{%
\pm}(k)-\mu }^{\prime }\,\mathbf{v}_{\pm}(\mathbf{k}) \mathbf{v}_{\pm}(%
\mathbf{k})}{\tilde{\omega}-\mathbf{v}_{\pm}(\mathbf{k})\cdot\mathbf{q}},
\label{eq1}
\end{equation}
where: (i) the band energies are linearized as $\varepsilon_{\pm}(k)=\pm
\hbar\mathrm{v}_{\text{\textsc{f}}} k$, with $\mathbf{v}_{\pm}(\mathbf{k})=%
\bm{\nabla}_{\mathbf{k}}\varepsilon_{\pm}(k)$ representing the electron
velocities and $\mathrm{v}_{\text{\textsc{f}}}=|\mathbf{v}_{\pm}(\mathbf{k}%
)| $ the Fermi velocity; (ii) $f_{\varepsilon _{\pm }(k)-\mu }^{\prime }$ is
the first energy-derivative of the FD, i.e.,
\begin{equation*}
f_{\varepsilon _{\pm }(k)-\mu }^{\prime } =\left. \frac{\partial
f_{\varepsilon -\mu }}{\partial \varepsilon }\right\vert _{\varepsilon
=\varepsilon _{\pm }(k)} =-\frac{1}{4kT\cosh ^{2}\left( \frac{\varepsilon
_{\pm }(k)-\mu }{kT}\right) };
\end{equation*}
(iii) $\tilde{\omega}=\omega- i\gamma$ is a complex frequency that includes
the angular frequency $\omega=2{\pi}f$ and a small shift along the imaginary
axes, which corresponds to the electron damping rate $\gamma$; the latter
depends on the average relaxation time $\tau$ as $\gamma =2\pi/\tau$.

In Eq.~\eqref{eq1}, the (two-dimensional)~first Brilluoin Zone~(BZ)
integration is well defined on circular areas of the $\mathbf{k}$-space
centered at the Dirac points, where the linear approximation for the band
energies is valid. Nevertheless, in most practical uses, an infinite
cone-structure is assumed for the valence and conduction bands; in other
words, the wave-vector integral is performed over the whole $\mathbf{k}$%
-space for analytical convenience and a factor of $2$ is included to account
for the inequivalent Dirac point. Then, a change of variable from
wave-vector to the energy leads to the non-dispersive conductivities $%
\bm{\sigma}_{\text{\textsc{rta}}}^{\pm }(\mathbf{0})=\sigma _{\mathrm{intra}%
}^{\pm }\mathbf{I}$, in which%
\begin{equation}
\sigma _{\mathrm{intra}}^{\pm }=\frac{ie^{2}\mathrm{v}_{\text{\textsc{f}}%
}^{2}}{\pi \tilde{\omega}}\int_{0}^{\infty }dkkf_{\varepsilon _{\pm }(k)-\mu
}^{\prime }=\frac{\pm ie^{2}}{\pi \hbar ^{2}\tilde{\omega}}\int_{0}^{\infty
}d\varepsilon \varepsilon f_{\varepsilon -\mu }^{\prime }  \notag
\end{equation}%
and $\mathbf{I}$ denotes the $2{\times }2$ identity matrix. The total
intra-band conductivity at $q\rightarrow 0$ is then%
\begin{eqnarray}
\sigma _{\mathrm{intra}} &=&\sigma _{\mathrm{intra}}^{+}+\sigma _{\mathrm{%
intra}}^{-}  \label{sigintra} \\
&=& \frac{ i e^{2}}{\pi \hbar^2 \tilde{\omega}}\int_{0}^{\infty}d\varepsilon
\varepsilon (f_{\varepsilon -\mu }^{\prime}-f_{-\varepsilon -\mu }^{\prime})
\notag
\end{eqnarray}%
Interestingly enough, the RTA conductivity~\eqref{eq1} and its $q\rightarrow
0$ form~\eqref{sigintra} can be derived from the Kubo formula in the optical
limit~\cite{12b,12c,12cc,12,12o}.

Another part of the graphene conductivity is related to inter-band processes
between the $\pi $ and $\pi ^{\ast }$ bands, and is also included in the
Kubo formulism~\cite{12c,12cc,12,12o}, though it does not contain the
dispersive term $\mathbf{v}_{\pm }(\mathbf{k})\cdot \mathbf{q}$. Indeed, it
has been pointed out that for surface waves supported by isolated graphene
sheets, and working frequencies below a few THz, the spatial dispersion
effects on inter-band transitions can be neglected~\cite{10,11}. Due to the
absence of the dispersive term, the inter-band conductivity is a scalar,
which can be expressed as
\begin{equation}
\sigma _{\mathrm{inter}}=\frac{ie^{2}}{\pi \hbar }\int_{0}^{\infty }\frac{%
d\varepsilon }{\hbar \tilde{\omega}}\frac{{f}_{\varepsilon -\mu }-{f}%
_{-\varepsilon -\mu }}{1-\left( \frac{2\varepsilon }{\hbar \tilde{\omega}}%
\right) ^{2}}.  \label{siginter}
\end{equation}%
Therefore, the total conductivity reads
\begin{equation}
\bm{\sigma}(\mathbf{q})=\bm{\sigma}_{\text{\textsc{rta}}}(\mathbf{q})+\sigma
_{\mathrm{inter}}\mathbf{I}.  \label{totsigrta}
\end{equation}%
In the $q\rightarrow 0$-limit the latter tends to the non-dispersive Kubo
conductivity~\cite{12c,12cc,12,12o}:
\begin{equation}
\sigma _{\text{\textsc{k}}}=\sigma _{\mathrm{intra}}+\sigma _{\mathrm{inter}%
}.  \label{eq:kubosig}
\end{equation}

Now, looking at the denominator in Eq.~\eqref{eq1}, it is clear that when $%
\mathbf{v}_{\pm}(\mathbf{k}) \cdot \mathbf{q}$ and $\omega$ are comparable,
spatial dispersion cannot be neglected in surface plasmon excitation. This
happens particularly in problems where a resonant behavior of the EM field
is concerned. It is the case of some of the examples reported in this work,
where the interaction between a radiating antenna and a graphene patch takes
place via near field coupling.

To give an idea of the numbers involved, the spatial harmonics of the
excitation field become significant in the conductivity response at an
operating frequency $f$ of the order of $\sim 1$~THz, for an applied
wave-vector value $q$ larger than $\sim 1~\mu$m$^{-1}$. At the same
frequency, the electric field wavelength is about $\sim 300~\mu$m. However,
the near field distribution between the tip of the antenna and the graphene
sample varies on a sub-micrometric scale, depending on the tip radius and
distance from the sample. Then, the slow-wave effect featured by plasmon
propagation implies a wavelength reduction of more than one order of
magnitude, and ensures the matching and coupling with the exciting near
field.

Eq.~(\ref{eq1}) can be simplified and made more explicit by expressing the
planar wave-vectors $\mathbf{q}$ and $\mathbf{k}$ in polar coordinates, say,
$\mathbf{q}=(q,\theta _{q})$ and $\mathbf{k}=(k,\vartheta _{k})$. Then, as
shown in the appendix, we can reduce it to the following expression
\begin{eqnarray}
\bm{\sigma}_{\text{\textsc{rta}}}^{\pm }(\mathbf{q}) &=&\frac{ie^{2}\mathrm{v%
}_{\text{\textsc{f}}}^{2}}{2\pi ^{2}}\int_{0}^{\infty }dkkf_{\varepsilon
_{\pm }(k)-\mu }^{\prime }  \label{eq3} \\
&&\qquad \times \int_{0}^{2\pi }d\vartheta _{k}\frac{\mathbf{I}+\left[
\begin{array}{cc}
\cos 2{\vartheta _{k}} & \sin 2{\vartheta _{k}} \\
\sin 2{\vartheta _{k}} & -\cos 2{\vartheta _{k}}%
\end{array}%
\right] }{\tilde{\omega}-\mathrm{v}_{\text{\textsc{f}}}q\cos (\vartheta
_{k}-\theta _{q})},  \notag
\end{eqnarray}%
where the BZ integral has been turned to a wave-vector integral over the
whole $\mathbf{k}$-space~(with the factor of $2$ from the inequivalent Dirac
points being included).

\begin{figure}[h]
\centerline{
\scalebox{0.99}{\includegraphics{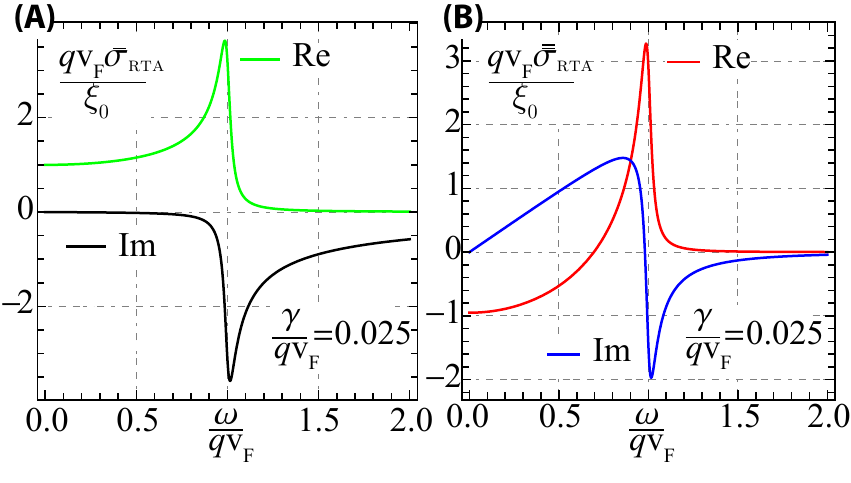}}
} \vskip -5pt
\caption{Real and imaginary parts of the scalar conductivities $\bar{\protect%
\sigma}_{\text{\textsc{rta}}}^{\pm }$ (A) and $\bar{\bar{\protect\sigma}}_{%
\text{\textsc{rta}}}^{\pm }$ (B) calculated from the RTA model~ \eqref{eq4},
and reported vs the dimensionless frequency $\protect\omega /q\mathrm{v}_{%
\text{\textsc{f}}}$. The two quantities are normalized to $\protect\xi _{0}/q%
\mathrm{v}_{\text{\textsc{f}}}$.}
\label{fig1A}
\end{figure}

After some straightforward manipulations on the angular integral in Eq.~%
\eqref{eq3} that are reported in the appendix, we obtain the tensor
components of the intra-band conductivity
\begin{equation}
\bm{\sigma}_{\text{\textsc{rta}}}^{\pm }(\mathbf{q})=\bar{\sigma}_{\text{%
\textsc{rta}}}^{\pm }(q)\mathbf{I}+\bar{\bar{\sigma}}_{\text{\textsc{rta}}%
}^{\pm }(q)\left[
\begin{array}{cc}
\cos 2{\theta _{q}} & \sin 2{\theta _{q}} \\
\sin 2{\theta _{q}} & -\cos 2{\theta _{q}}%
\end{array}%
\right] ,  \label{eq4}
\end{equation}%
which depend on the scalar conductivities
\begin{eqnarray*}
\bar{\sigma}_{\text{\textsc{rta}}}^{\pm }(q) &=&\frac{ie^{2}\mathrm{v}_{%
\text{\textsc{f}}}^{2}}{2\pi ^{2}\tilde{\omega}}\int_{0}^{\infty
}dkkf_{\varepsilon _{k}^{\pm }-\mu }^{\prime } \\
&&\qquad \qquad \times \int_{0}^{2\pi }\frac{d\theta _{k}}{\tilde{\omega}-(%
\mathrm{v}_{\text{\textsc{f}}}q/\tilde{\omega})\cos \theta _{k}}
\end{eqnarray*}%
and
\begin{eqnarray*}
\bar{\bar{\sigma}}_{\text{\textsc{rta}}}^{\pm }(q) &=&\frac{ie^{2}\mathrm{v}%
_{\text{\textsc{f}}}^{2}}{2\pi ^{2}\tilde{\omega}}\int_{0}^{\infty
}kf_{\varepsilon _{k}^{\pm }-\mu }^{\prime } \\
&&\qquad \qquad \times \int_{0}^{2\pi }\frac{d\theta _{k}\cos 2\theta _{k}}{%
\tilde{\omega}-(\mathrm{v}_{\text{\textsc{f}}}q/\tilde{\omega})\cos \theta
_{k}}.
\end{eqnarray*}%
These two quantites may rewritten as%
\begin{equation*}
\bar{\sigma}_{\text{\textsc{rta}}}^{\pm }(q)=\frac{\sigma _{\mathrm{intra}%
}^{\pm }}{\sqrt{1-\mathrm{v}_{\text{\textsc{f}}}^{2}q^{2}/\tilde{\omega}^{2}}%
}
\end{equation*}%
and%
\begin{equation*}
\bar{\bar{\sigma}}_{\text{\textsc{rta}}}^{\pm }(q)=\frac{\left( \sqrt{1-%
\mathrm{v}_{\text{\textsc{f}}}^{2}q^{2}/\tilde{\omega}^{2}}-1\right) ^{2}}{%
\sqrt{1-\mathrm{v}_{\text{\textsc{f}}}^{2}q^{2}/\tilde{\omega}^{2}}}\frac{%
\tilde{\omega}^{2}\sigma _{\mathrm{intra}}^{\pm }}{\mathrm{v}_{\text{\textsc{%
f}}}^{2}q^{2}}.
\end{equation*}%
Therefore, summing over the $\pm $ channels, we find
\begin{equation*}
\bar{\sigma}_{\text{\textsc{rta}}}(q)=\bar{\sigma}_{\text{\textsc{rta}}%
}^{+}(q)+\bar{\sigma}_{\text{\textsc{rta}}}^{-}(q)=\frac{\sigma _{\mathrm{%
intra}}}{\sqrt{1-\mathrm{v}_{\text{\textsc{f}}}^{2}q^{2}/\tilde{\omega}^{2}}}
\end{equation*}%
and
\begin{eqnarray*}
\bar{\bar{\sigma}}_{\text{\textsc{rta}}}(q) &=&\bar{\bar{\sigma}}_{\text{%
\textsc{rta}}}^{+}(q)+\bar{\sigma}_{\text{\textsc{rta}}}^{-}(q) \\
&=&\frac{\left( \sqrt{1-\mathrm{v}_{\text{\textsc{f}}}^{2}q^{2}/\tilde{\omega%
}^{2}}-1\right) ^{2}}{\sqrt{1-\mathrm{v}_{\text{\textsc{f}}}^{2}q^{2}/\tilde{%
\omega}^{2}}}\frac{\tilde{\omega}^{2}\sigma _{\mathrm{intra}}}{\mathrm{v}_{%
\text{\textsc{f}}}q^{2}}.
\end{eqnarray*}%
Then, $\bar{\sigma}_{\text{\textsc{rta}}}$ and $\bar{\bar{\sigma}}_{\text{%
\textsc{rta}}}$ turn out to be both proportional to the non-dispersive
intra-band conductivity $\sigma _{\mathrm{intra}}$ of Eq.~\eqref{sigintra},
which evaluates exactly to $\sigma _{\mathrm{intra}}=-i\xi _{0}/\tilde{\omega%
}$ with
\begin{equation*}
\xi _{0}=\frac{e^{2}}{\pi \hbar ^{2}}[\mu +2kT\ln (e^{-\frac{\mu }{kT}}+1)].
\end{equation*}%
Incidentally, we notice that the conductivity $\bar{\bar{\sigma}}_{\text{%
\textsc{rta}}}^{\pm }$, being markedly anisotropic, is proportional to $%
\tilde{\omega}^{2}\sigma _{\mathrm{intra}}$. This makes its imaginary part
change sign with increasing the frequency, in contrast to the purely
inductive nature of $\bar{\sigma}_{\text{\textsc{rta}}}^{\pm }$, whose
imaginary part keeps a negative sign in the sampled frequency range. The
plots of Fig.~\ref{fig1A}\textbf{(A)} and~\ref{fig1A}\textbf{(B)} illustrate
such a behavior. Now, by definition, surface waves are sustained at the
interface between materials having different permittivities so that, in a
transverse resonance circuit, their associated reactances cancel out.
Therefore, the contribution of the anisotropic term $\bar{\bar{\sigma}}_{%
\text{\textsc{rta}}}^{\pm }$ may strongly affect the polarization of surface
plasmons~\cite{14} of high rate of spatial variation.

To include the diffusion currents, we introduce the following tensor
quantity
\begin{equation}
\mathbf{s}_{\pm }(\mathbf{q})=\frac{- i\gamma}{\omega}\int\limits_{1^{%
\mathrm{st}}~\mathrm{BZ}}d^{2}k \frac{\tilde{f}_{\varepsilon_{\pm}(k)-\mu
}^{\prime }\,\mathbf{v}_{\pm }(\mathbf{k})\mathbf{q}}{\tilde{\omega}-\mathbf{%
v}_{\pm }(\mathbf{k})\cdot \mathbf{q}},
\end{equation}
where
\begin{equation*}
\tilde{f}_{\varepsilon_{\pm}(k)-\mu }^{\prime }=\frac{f_{\varepsilon_{%
\pm}(k)-\mu }^{\prime }}{\int\limits_{1^{\mathrm{st}}~\mathrm{BZ}%
}d^{2}kf_{\varepsilon_{\pm}(k)-\mu }^{\prime }}.
\end{equation*}
The BGK correction to the RTA conductivity can be put in the form
\begin{equation}  \label{eq4b}
\bm{\sigma}_{\text{\textsc{bgk}}}^{\pm}(\mathbf{q})=[\mathbf{I}+\mathbf{s}%
_{\pm }(\mathbf{q})]^{-1}\bm{\sigma}_{\text{\textsc{rta}}}^{\pm}(\mathbf{q}).
\end{equation}
This is equivalent to the result presented in Ref.~\cite{10}, and leads to
correct the total conductivity as
\begin{equation}
\bm{\sigma}(\mathbf{q})=\bm{\sigma}_{\text{\textsc{bgk}}}(\mathbf{q}%
)+\sigma_{\mathrm{inter}}\mathbf{I}.  \label{totsigbgk}
\end{equation}

The numerical results reported in the following are obtained by both a
full-wave solver and a semi-analytical approach, as implemented respectively
by the finite element method~(FEM) and the method of moments~(MoM). The
latter employs the usual free-space Green's tensor $\mathbf{G}$, in
cylindrical coordinates $\mathbf{r}=(r,\varphi)$ and $z$, as the kernel of
an integral operator relating the electric current density and field~\cite%
{12,12o}:
\begin{equation*}
\mathbf{E}(\mathbf{r},z)= \int d^2 r^{\prime} \int dz^{\prime} \mathbf{G}(%
\mathbf{r}-\mathbf{r}^{\prime},z-z^{\prime })\cdot \mathbf{J}(\mathbf{r}%
^{\prime },z^{\prime }).
\end{equation*}
Here, it should be noticed that the discretized currents in space are
assumed as independent variables in the MoM calculation. Then, the solution
for the resulting Electric Field Integral Equation~(EFIE) is generalized to
the case of spatially dispersive material~\cite{11,12,12o}, by using the
proper constitutive relation in the spatial domain, i.e.,
\begin{equation}  \label{eq2}
\mathbf{J}_{\text{\textsc{s}}}(\mathbf{r})= \int d^2 r^{\prime} \bm{\sigma}(%
\mathbf{r}-\mathbf{r}^{\prime }) \cdot \mathbf{E}_{\text{\textsc{s}}}(%
\mathbf{r}^{\prime }).
\end{equation}
In this equation, the planar current $\mathbf{J}_{\text{\textsc{s}}}$, and
the tangent electric field $\mathbf{E}_{\text{\textsc{s}}}$, are sampled
over the graphene surface.

The non-local behavior of the current-field relation becomes important when
rapid field variations are involved. Transforming the conductivity in real
space, the input spatial response of dispersive graphene is obtained as
\begin{eqnarray}
\bm{\sigma}({\mathbf{r}}) &=&\mathbf{I}\int_{0}^{\infty }\frac{dq}{2\pi }q%
\bar{\sigma}(q)J_{0}(qr)+\mathbf{I}\sigma _{\mathrm{inter}}  \label{eq5} \\
&&-\left(
\begin{array}{cc}
\cos 2\phi & \sin 2\phi \\
\sin 2\phi & -\cos 2\phi%
\end{array}%
\right) \int_{0}^{\infty }\frac{dq}{2\pi }q\bar{\bar{\sigma}}(q)J_{2}(qr),
\notag
\end{eqnarray}%
where the scalar conductivities $\bar{\sigma}$ and $\bar{\bar{\sigma}}$
include both the $\pm $-contributions from Eq.~\eqref{eq4} or Eq.~%
\eqref{eq4b}. In Eq.~\eqref{eq5} the first and second terms contributing to
the conductance, are weighted by the zero-order and second-order Bessel
functions, respectively. This means that the first addend of the
conductivity is more sensitive to slow varying fields with respect to the
second addend. Equally importantly, the second term of the conductivity,
containing off-diagonal matrix elements, is characterized by an angular
dependence related to a current response, which is locally weighted by a
\textquotedblleft quadrupole\textquotedblright\ spatial distribution of the
EM field. Consequently, unless very high angular and radial variations of
the EM fields are concerned, the second term can be neglected.

\begin{figure}[!h]
\centerline{
\scalebox{0.99}{\includegraphics{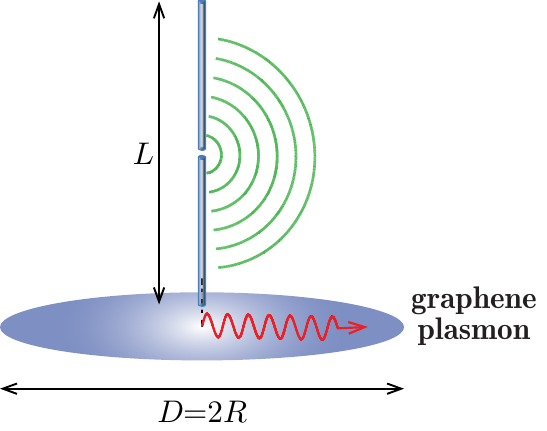}}} \vskip -5pt
\caption{Wire antenna placed at sub-micrometric distance~($L/200$) from a
graphene patch}
\label{fig0}
\end{figure}

Let us now turn to the main application of the present work~(Fig.~\ref{fig0}%
) that is a circular graphene disk of diameter $D=2R$, with a wire antenna
of length $L$, placed just above its center. For this system, we can safely
assume cylindrical symmetry. The gap between the antenna and the disk is $%
L/200$. In absence of angular variation of the EM excitation, the off
diagonal terms vanish and no angular current arises. In addition, with a not
too small tip-sample distance, and, thus, a not too strong EM field
variation, the $\bar{\sigma}$-term in Eq.~\eqref{eq5} is expected to be
dominant. Under these limiting conditions, Eq.~\eqref{eq2} can be
approximated by
\begin{equation}  \label{eq7}
J_r = \int d^2r^{\prime }\bar{\sigma}(|\mathbf{r}-\mathbf{r}^{\prime }|) E_{%
\text{\textsc{s}}}(\mathbf{r}^{\prime })
\end{equation}
where $J_r$ is the radial component of the surface current, and $\bar{\sigma}%
(r)$ is the real space representation of $\bar{\sigma }(q )$, which gives
the (spatial)~impulsive response of the current after the EM excitation.
Note that usual assumption of a thin hollow cylinder to approximate the wire
antenna may affect the near field distribution between the tip and graphene,
but it does not limit the generality of the present analysis.

\begin{figure}[h]
\centerline{
\scalebox{0.99}{\includegraphics{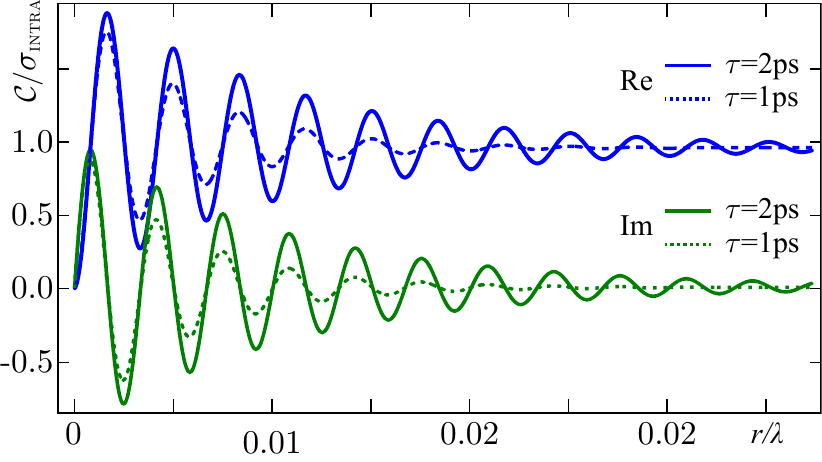}}
} \vskip -5pt \vskip -5pt
\caption{ Cumulative integral of the impulsive response $\mathcal{C}/\protect%
\sigma_{\mathrm{intra}}$ as a function of radial position $r$, normalized to
the free-space wavelength $\protect\lambda $; two different scattering
times~($\protect\tau =1,2$~ps) are tested.}
\label{fig1}
\end{figure}

For a fixed frequency, the effect of spatial dispersion increases with
increasing the charge scattering-time.
As a practical example, we take the cumulative integral of the impulsive
response
\begin{equation*}
\mathcal{C}(r)=\int\limits_{r^{\prime }<r}d^{2}r^{\prime }\bar{\sigma}%
(r^{\prime }) =2\pi \int_{0}^{r}dr^{\prime }r^{\prime }\bar{\sigma}%
(r^{\prime}),
\end{equation*}
i.e., the current density response to a uniform unit electric field within a
circular area of radius $r<R$. Its profile, normalized to $\sigma_{\mathrm{%
intra}}$, is shown in Fig.~\ref{fig1} for two different scattering times~($%
\tau =1,2$~ps), at $f=10$~THz.

In absence of dispersion, the response of Fig.~\ref{fig1} would be a real
constant without spatial ripples. In presence of dispersion, the actual
response is more complex, and some spatial ripples appear. These
oscillations extend just to a fraction of the free-space wavelength $\lambda$%
, which is typically comparable with the plasmon wavelength, and increases
with increasing the charge lifetime. We expect that the above behavior of
the system will reflect, numerically, on the solution of the EM field
distribution.

\section{Numerical Results}

\label{sec:numres} In the following, we present the solution for the EFIE
directly in real space, focussing on both the full tensor form~\eqref{eq5}
and the scalar form~\eqref{eq7} of the surface conductivity.

\subsection{Non-Dispersive Analysis of Surface Plasmons}

\label{sub:nondisppla} We begin by considering an example of plasmon
excitation without spatial dispersion. Let us define the graphene surface
impedance $Z_{\text{\textsc{s}}}$ as the reciprocal of the non-dispersive
Kubo conductivity $\sigma_{\text{\textsc{k}}}$ introduced in Sec.~\ref%
{sec:theory}. We then have $Z_{\text{\textsc{s}}}=1/\sigma_{\text{\textsc{k}}%
}$ or $1/Z_{\text{\textsc{s}}}=\sigma_{\mathrm{intra}}+\sigma_{\mathrm{inter}%
}$, where the intra-band and inter-band terms have been respectively given
in Eqs.~\eqref{sigintra} and~\eqref{siginter}.
\begin{figure}[h]
\centerline{
\scalebox{0.99}{\includegraphics{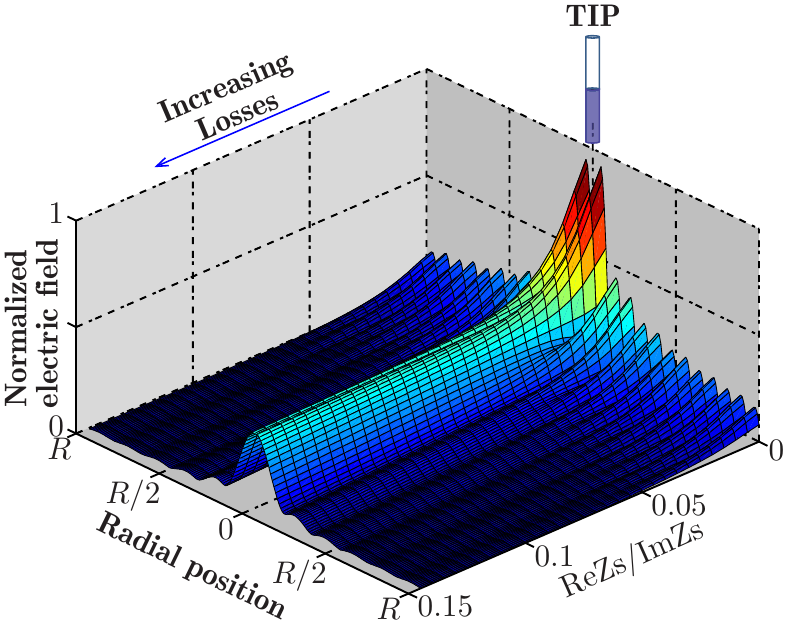}}
} \vskip -5pt
\caption{Plasmon distribution~(normalized electric field) from a linear
antenna coupled to a graphene disk~(sketched in Fig.~\protect\ref{fig0}) vs
the radial distance from the disk center and $\mathrm{Re}{Z_{\text{\textsc{s}%
}}}/\mathrm{Im}{Z_{\text{\textsc{s}}}}$.}
\label{fig2}
\end{figure}

The plasmon distribution for a graphene disk of diameter $D$, coupled to an
antenna of length $L=D\approx c/2f$, is reported in Fig.~\ref{fig2} vs the
radial position within the disk and the inverse plasmon ``\textit{quality
factor}". The latter is defined as the ratio between imaginary and real
parts of $Z_{\text{\textsc{s}}}$ calculated by the non-dispersive Kubo
conductivity, i.e., $\mathrm{Im}{Z_{\text{\textsc{s}}}}/\mathrm{Re}{Z_{\text{%
\textsc{s}}}}=-\mathrm{Re}{\sigma_{\text{\textsc{k}}}}/\mathrm{Im}{\sigma_{%
\text{\textsc{k}}}}$. In Fig.~\ref{fig2}, both the operation frequency and
the graphene chemical potential follow from the choice of the surface
reactance $\mathrm{Im}{Z_{\text{\textsc{s}}}}$ and plasmon quality-factor.
For example, a reactance of $5$~k$\Omega $ and a quality factor of $11$ are
associated to a chemical potential of $0.08$~eV and a frequency of about $%
7.4 $~THz. We see that the spatial oscillations and propagation of the
plasmon expire, at a progressively smaller radial distance from the feeding
tip, as $\mathrm{Re}{Z_{\text{\textsc{s}}}}/\mathrm{Im}{Z_{\text{\textsc{s}}}%
}$ increases from $0$ to $0.15$.

\begin{figure}[h]
\centerline{\scalebox{0.925}{\includegraphics{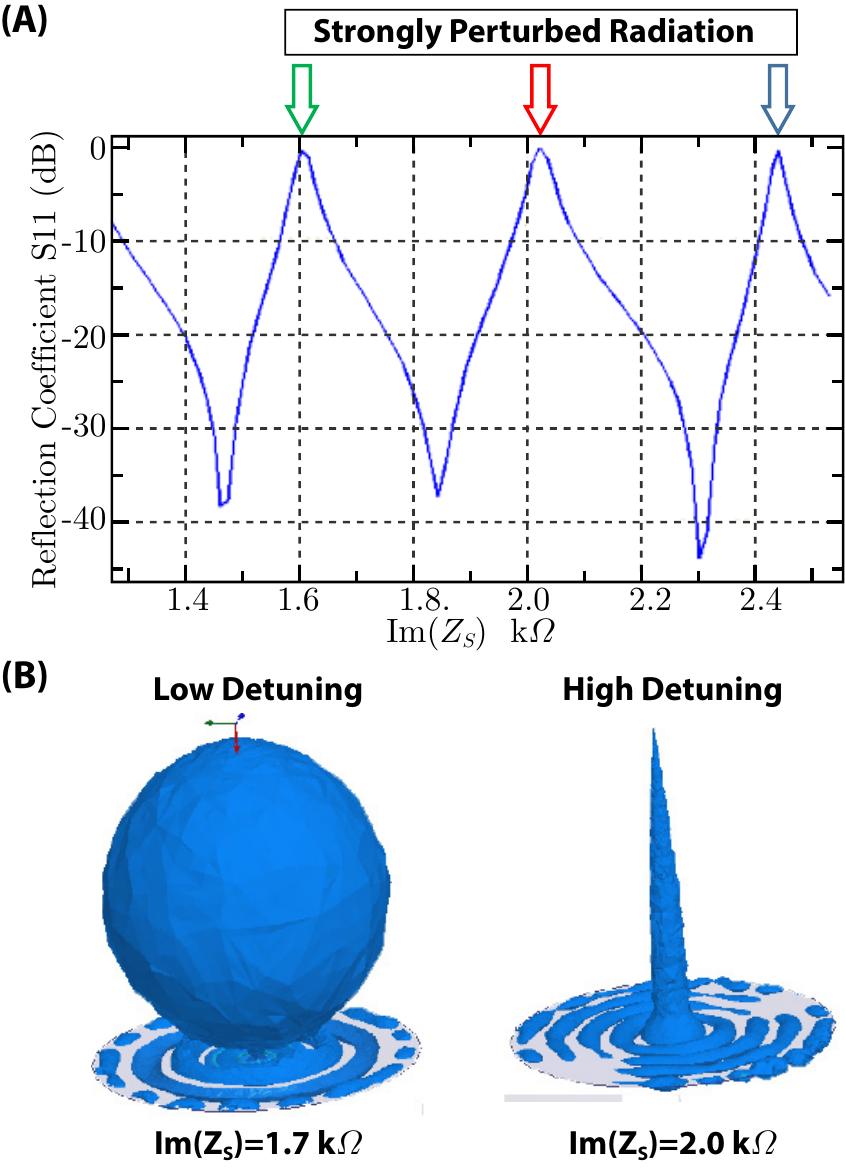}}} \vskip -5pt
\caption{ (A) Input matching of the antenna vs surface reactance; (B)
iso-surface plot of the electric field (magnitude) radiated by the wire
antenna, and plasmon excitation on the underlying graphene patch.}
\label{fig3}
\end{figure}
Another interesting effect is pointed out in Fig.~\ref{fig3}, where the
overlap of the plasmon distribution with the near field under the tip
determines the strength of the coupling and the amount of power transferred
from the antenna to the plasmon. In order to emphasize this concept- and
make the effect more evident- the losses are just neglected. More
specifically, the surface resistivity is taken to be an independent
variable, i.e., its value is not calculated from the Kubo conductivity as in
Fig.~\ref{fig2}. In addition, $Z_{\text{\textsc{s}}}$ is considered a purely
imaginary quantity~(no losses) and the reactance $\mathrm{Im}{Z_{\text{%
\textsc{s}}}}$ is varied from about $\sim 1.3$~k$\Omega $ to $\sim 2.5$~k$%
\Omega$. Thus, the geometric parameters may be expressed in terms of the
sizes of the antenna and the graphene disk relative to the vacuum wavelength
$\lambda$. In particular: (i) the diameter $D$ of the patch and the length $%
L $ of the wire antenna are related by $D=L=2.11c/f=2.11\lambda$; (ii) the
air-gap distance $d$ between the antenna and the graphene-patch is given by $%
d=L/200$; (iii) the radius $a$ of the antenna is fixed to $a=L/1000$; (iv) a
value of $L/200$ is chosen for the \textit{excitation-gap} $g$ in the middle
of the antenna, where the \textit{voltage-source} is applied; (v) The
internal impedance $Z_0$ of the voltage applied to the excitation-gap of the
antenna is set to $70~\Omega$.

It should be noted that the value of $g$ can be scaled up or down provided
that it remains much smaller than $\lambda$ and that the applied voltage is
scaled reversely. In addition, the internal impedance value is equal to the
real part of the input impedance $Z11$ of the antenna \textit{in absence of
graphene} to have maximum power transfer. Fig.~\ref{fig3}(\textbf{A}) shows
that the antenna, which would work precisely at its resonance frequency in
absence of graphene, can be more or less ``detuned'' by the graphene patch.
The detuning depends on the graphene-antenna coupling: the higher the
coupling strength, the higher the coupled reactive power, and the higher the
reflection coefficient $S11$ at the input port of the antenna~(indicated by
arrows). $S11$ expresses the amount of power that is reflected back at the
terminals of a voltage-source excitation, given by an infinitesimal electric
dipole located at the center of the antenna. By definition, we have $%
S11=(Z11-Z_0)/(Z11+Z_0)$. The choice of the source impedance of the voltage
excitation is arbitrary. However, as stated above, we have set $Z_0$ in
order to have resonance in absence of graphene, and accordingly a better
visualization of the perturbing effect of the graphene patch~(see Sec.~\ref%
{sub:disppla} for more information and numerical values). Fig.~\ref{fig3}(%
\textbf{B}) shows in details what is happening to the antenna near~(right
picture) and far from~(left picture) the detuning points, i.e., the arrows
of Fig.~\ref{fig3}(\textbf{A}). In particular, we have taken surface
reactance values of $2$ and $1.7$~k$\Omega$, corresponding respectively to
input matching values of about $-4$ and $-13$dB: in the former case, the
radiation is drastically reduced by the plasmon coupling and a large amount
of power is reflected back to the input port of the antenna. The results
reported in Fig.~\ref{fig3}(\textbf{A}) are derived from our MoM simulator~%
\cite{15}, whereas the plots of Fig.~\ref{fig3}(\textbf{B}) are calculated
by a full-wave EM solver~(HFSS by Ansoft), which provides an independent
validation of our implementation.

\subsection{Dispersive Characterization of Surface Plasmons}

\label{sub:disppla} 
\begin{figure}[h]
\centerline{
\scalebox{0.95}{\includegraphics{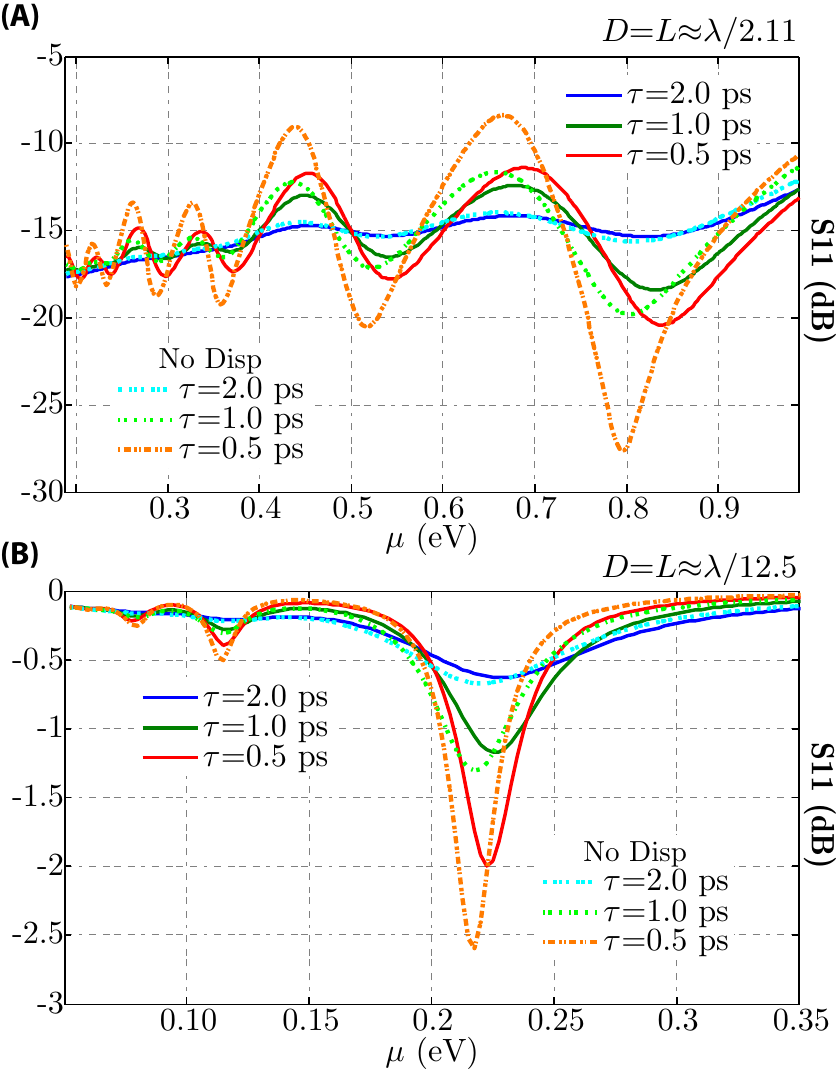}}} \vskip -5pt
\caption{ Reflection coefficient at the input terminal of the antenna
with~(solid lines) and without~(dashed lines) spatial dispersion, for
different relaxation times. The operating frequency is set to $10$ THz. Two
disk diameters are considered, namely $D=L\approx \protect\lambda/2.11$ (A),
$D=L\approx \protect\lambda/12.5$ (B). }
\label{fig4}
\end{figure}

In order to discuss the effect of dispersion on the
strength of plasmon excitation, we focus on the two examples reported in
Fig.~\ref{fig4}. Differently from the results shown in Fig.~\ref{fig3}, here
we take into account ohmic losses using the approximate complex conductivity
of Eq.~\eqref{eq7} within the RTA limit. Assuming a nominal frequency of $10$%
~THz, we show the effect of spatial dispersion on the input matching~(S11)
of the antenna in resonant~[$L\approx \lambda/2.11$, Fig.~\ref{fig4}\textbf{%
(A)}] and non-resonant~[$L\approx \lambda/12.5$, Fig.~\ref{fig4}\textbf{(B)}%
] conditions, respectively. All the reflection coefficients at the input
terminal of the antenna are plotted as function of the chemical potential of
graphene, i.e., the Fermi energy shift associated to the local doping level.

Three typical values of charge-carrier relaxation times on the ps time-scale
are tested~($\tau=0.5,1,2$~ps). Indeed, the relaxation time in graphene may
strongly depend on the quality of the sample related to the fabrication
process. In particular, real samples are affected by the presence of grain
boundaries, defects, multilayer regions, etc. For the above reason, some
flexibility is needed in selecting the $\tau$ parameter. As evident from
Fig.~\ref{fig4}, the larger is the relaxation time the sharper are the S11
peaks that express the maxima and minima of the antenna-plasmon coupling.
The antenna is fed by a voltage source with an impedance equal to the real
part of the input impedance of the antenna, which is about $70~\Omega $ and $%
2222~\Omega $ for the resonant and non-resonant cases, respectively.
Clearly, the minima of reflection, corresponding to surface field
resonances, have lower matching levels in the non-resonant case. In either
resonant or non resonant condition, the dispersive reflection peaks are more
broadened- and have lower maxima- with respect to the corresponding
non-dispersive ones. These differences become more and more evident as the
relaxation time increases, or, the quality of graphene gets better.

\begin{figure}[h]
\centerline{
\scalebox{0.99}{\includegraphics{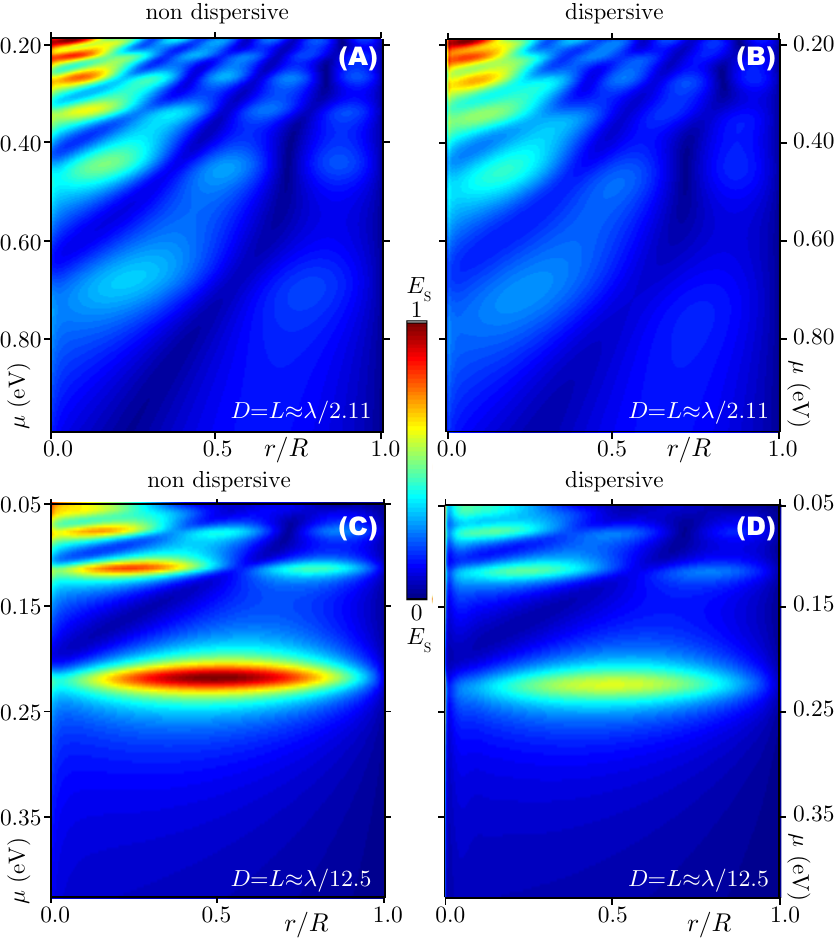}}} \vskip -5pt
\caption{ Spatial distribution of the surface electric field $E_s$%
~(normalized to $1$) as function of relative position~($r/R$) and chemical
potential~$\protect\mu$, with (B,D) and without (A,C) spatial dispersion
effects, for f=10~THz, and $D=L\approx \protect\lambda/2.11$ (A,B), $%
D=L\approx \protect\lambda/12.5$ (C,D)}
\label{fig5}
\end{figure}

In SMM, an electrostatic tuning of the charge density with a DC voltage
applied to the microscope tip~\cite{7} can provide the local doping for near
field applications. Fig.~\ref{fig5} shows the spatial distributions of the
plasmonic wave on the graphene disk~(for a normalized electric field) as
function of the radial position and the doping levels~(values corresponding
to those of Fig.~\ref{fig4}). The distributions are computed with and
without spatial dispersion, assuming a relaxation time of $2$~ps.
Differences between the dispersing and non dispersing curves are clearly
observable, particularly in correspondence of the resonance peaks. Looking
at the plots of Fig.~\ref{fig4} and Fig.~\ref{fig5}, we see that the
absolute effect of spatial dispersion is higher in the non-resonant case,
where the size of the graphene disk is much smaller than the free-space
wavelength $\lambda$.

\begin{figure}[h]
\centerline{
\scalebox{0.99}{\includegraphics{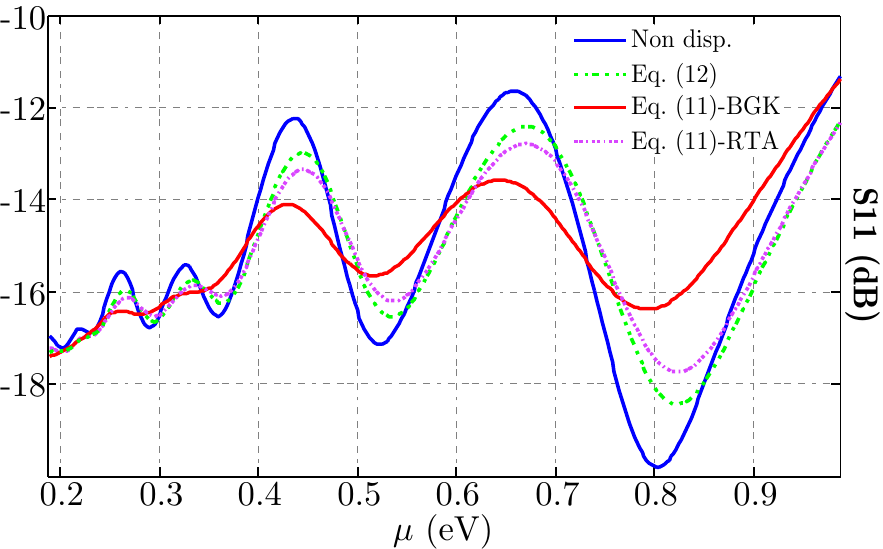}}} \vskip -5pt
\caption{ Reflection coefficient~(S11) at input terminal of the antenna with
the different approaches to dispersion discussed here. The following
parameters are used: $f=10$~THz, $\protect\tau=1$~ps, and $D=L\approx\protect%
\lambda/2.11$.}
\label{fig6}
\end{figure}

To go beyond the diagonal conductivity approximation of Eq.~\eqref{eq7} in
the RTA, we explicitly account for the angular variations of the EM field
and, at the same time, we include the effect of diffusion currents. Thus, we
next consider the full conductivity response of Eq.~\eqref{eq5}, within both
the RTA and BGK approach. In Fig~\ref{fig6}, we report the details of the
input matching of the antenna in the resonant case $D\approx L=\lambda/2.12$%
, comparing the results obtained from the different levels of approximations
discussed in the present work. We find confirmation that the RTA
conductivity of the graphene disk represents a significant improvement with
respect to the non-dispersive conductivity of an infinite graphene sheet. In
addition, we see that the contribution of the off diagonal tensor $\bar{\bar{%
\sigma}}$ leads to a non-negligible small correction to the diagonal tensor $%
\bar{\sigma}$, within the RTA.

More importantly, we notice that the effect of electron diffusion included
in the BGK model plays a significant role. Indeed the RTA and BGK
expressions of $\bar{\sigma}$ and $\bar {\bar {\sigma }}$ are remarkably
different: dispersion effects appear to be under-estimated by the RTA
conductivity~[or even more simplified expressions as Eq.~\eqref{eq7}] with
respect to the BGK form.

\section{Conclusions}

We have used a semiclassical model derived from the Boltzmann transport
equation to investigate the effect of spatial dispersion on the linear THz
response of graphene, characterized by excitation of surface plasmons. The
excitation source has been provided by the near field of an antenna
radiating in proximity of a graphene micro-disk.

We have characterized the role of spatial dispersion, with respect to a
cylindrical system, within both the RTA and the BGK approach, obtaining
meaningful and compact expressions of the full-tensor representing the
constitutive relation of the graphene patch in real space.

We have shown that the surface distribution of the field on the disk, and
the macroscopic response of the antenna, is significantly affected by
spatial dispersion in two distinct noteworthy examples, where the antenna
has been set in resonant and non-resonant conditions.

Although the role played by spatial dispersion was previously clarified in
propagation and modal analysis of plasmons~\cite{10,11}, here we have
provided a focus on near field excitation problems, with potential fallout
in important practical applications, concerning near field imaging.
\Allow{
\begin{acknowledgments}
This work has been supported by the European Project ``\textit{Carbon Based Smart Systems for Wireless Applications}'', (NANO-RF, n. 318352). We also thank NOW Srli (www.notonlywaves.com) for providing HFSS simulations.
\end{acknowledgments}

\begin{appendix}
}

\section*{Appendix: Dispersive and Non-Dispersive RTA conductivity}

Using the polar coordinates $\mathbf{k}=k(\cos \vartheta _{k},sin\vartheta
_{k})$ and $\mathbf{q}=q(\cos \theta _{q},\sin \theta _{q})$, the tensor
product at the numerator of the BZ integral in Eq.~\eqref{eq1} becomes
\begin{eqnarray*}
\mathbf{v}_{\pm }(\mathbf{k})\mathbf{v}_{\pm }(\mathbf{k}) &=&\mathrm{v}_{%
\text{\textsc{f}}}^{2}\left[
\begin{array}{cc}
\cos ^{2}\vartheta _{k} & \sin \vartheta _{k}\cos \vartheta _{k} \\
\sin \vartheta _{k}\cos \vartheta _{k} & \sin ^{2}\vartheta _{k}%
\end{array}%
\right]  \\
&=&\frac{\mathrm{v}_{\text{\textsc{f}}}^{2}}{2}\left\{ \mathbf{I}+\left[
\begin{array}{cc}
\cos 2{\vartheta _{k}} & \sin 2{\vartheta _{k}} \\
\sin 2{\vartheta _{k}} & -\cos 2{\vartheta _{k}}%
\end{array}%
\right] \right\} ,
\end{eqnarray*}%
while the scalar products at the denominator reads: $\mathbf{v}_{\pm }(%
\mathbf{k})\cdot \mathbf{q}=\mathrm{v}_{\text{\textsc{f}}}q\cos (\vartheta
_{k}-\theta _{q})$. Hence, Eq.~\ref{eq1} is turned to Eq.~\ref{eq3}. Now,
consider the integral identity
\begin{equation*}
\int_{0}^{2\pi }\frac{d\psi \cos n\psi }{1+z\cos \psi }=\frac{2\pi (-1)^{n}}{%
\sqrt{1-z^{2}}}\left( \frac{\sqrt{1-z^{2}}-1}{z}\right) ^{n},
\end{equation*}%
which holds true for $n=0,1,\ldots $ and $\mathrm{Im}{z}\neq 0$, and can be
derived from the database~\cite{13}. As special cases, we get
\begin{equation*}
A(z)=\int_{0}^{2\pi }\frac{d\psi }{1-z\cos \psi }=\frac{2\pi }{\sqrt{1-z^{2}}%
}
\end{equation*}%
for $n=0$ and
\begin{equation*}
B(z)=\int_{0}^{2\pi }\frac{d\psi \cos 2\psi }{1-z\cos \psi }=\frac{2\pi }{%
\sqrt{1-z^{2}}}\left( \frac{\sqrt{1-z^{2}}-1}{z}\right) ^{2}
\end{equation*}%
for $n=2$. By a simple change of variable, we also have:
\begin{eqnarray*}
\int_{0}^{2\pi }\frac{d\psi \cos 2\psi }{1-z\cos (\psi -\psi _{0})}
&=&B(z)\cos 2\psi _{0}, \\
\int_{0}^{2\pi }\frac{d\psi \sin 2\psi }{1-z\cos (\psi -\psi _{0})}
&=&B(z)\sin 2\psi _{0}.
\end{eqnarray*}%
It follows that the $\vartheta _{k}$-integral in Eq.~(\ref{eq3}) evaluates
to
\begin{multline*}
\int_{0}^{2\pi }d\vartheta _{k}\frac{\mathbf{I}+\left[
\begin{array}{cc}
\cos 2{\vartheta _{k}} & \sin 2{\vartheta _{k}} \\
\sin 2{\vartheta _{k}} & -\cos 2{\vartheta _{k}}%
\end{array}%
\right] }{\tilde{\omega}-\mathrm{v}_{\text{\textsc{f}}}q\cos (\vartheta
_{k}-\theta _{q})}=\frac{A(\mathrm{v}_{\text{\textsc{f}}}q/\tilde{\omega})}{%
\tilde{\omega}}\mathbf{I} \\
+\frac{B(\mathrm{v}_{\text{\textsc{f}}}q/\tilde{\omega})}{\tilde{\omega}}%
\left[
\begin{array}{cc}
\cos 2\theta _{q} & \sin 2\theta _{q} \\
\sin 2\theta _{q} & -\cos 2\theta _{q}%
\end{array}%
\right] .
\end{multline*}%
Plugging this in to Eq.~(\ref{eq3}), we obtain Eq.~(\ref{eq4}) and the
scalar conductivities $\bar{\sigma}_{\mathrm{rta}}^{\pm }$, $\bar{\bar{\sigma%
}}_{\mathrm{rta}}^{\pm }$. \Allow{\end{appendix}}

\end{document}